Technical Note

# A Gap between Simulation and Practice for Recursive Filters:

# On the State Transition Noise

Tiancheng Li

*Abstract* In order to evaluate and compare different recursive filters, simulation is a common tool and numerous simulation models are widely used as 'benchmark'. In the simulation, the continuous time dynamic system is converted into a discrete-time recursive system. As a result of this, the state indeed evolves by Markov transitions in the simulation rather than in continuous time. One significant issue involved with modeling of the system from practice to simulation is that the simulation parameter, particularly e.g. the state Markov transition noise, needs to match the iteration period of the filter. Otherwise, the simulation performance may be far from the truth. Unfortunately, quite commonly different-speed filters are evaluated and compared under the same simulation model with the same state transition noise for simplicity regardless of their real sampling periods. Here the note primarily aims at clarifying this problem and point out that it is very necessary to use a proper simulation noise that matches the filter's speed for evaluation and comparison under the same simulation model.

T. Li's research interest is nonlinear filtering (especially particle filters), multi-target tracking, information fusion, artificial intelligence. He is currently with the School of Mechatronics, Northwestern Polytechnical University, Xi'an, 710072, China. He is also with the Center for Automated and Robotics NDT, London South Bank University, London, SE1 0AA, UK
Contact email: lit3@lsbu.ac.uk, robottcli@gmail.com

Filtering estimation is a class of signal processing that widely exists in engineering and therefore is a very broad research topic. The solution of the continuous time filtering problem can be represented as a ratio of two expectations of certain functionals of the signal process [1, 2]. These functionals are parametrized by the observation path $\{y_s, s \geq 0\}$. However, in practice, only the values of the observation corresponding to a discrete time partition are available, i.e., $\{y_k, k = 0, 1, \ldots\}$. Also for the reason of flash-frequency-based computing on the computer, the continuous-time dynamic system has to be converted into a discrete-time simulation model e.g. discrete Markov System by discretely sampling the outputs, namely discretization. In which, the recursive iteration duration is just the sampling period of the filter. The note considers the time-varying filtering problem in the discrete dynamic State Space Model (SSM)

$$\begin{cases} x_k = f_k(x_{k-1}, u_{k-1}, w_{k-1}) & \text{(state dynamic equation)} \\ y_k = h_k(x_k, v_k) & \text{(measurement equation)} \end{cases} \quad (1)$$

where $k$ indicates the discrete time-step (also called as frames) which can only be a positive integer, $x_k$ denotes the state vector, $y_k$ denotes the measurement vector, $u_k$ denotes the system input vector (known as driving force in a controlled environment), $w_k$ and $v_k$ denote stochastic noise with appropriate dimensions affecting the Markov dynamic $f_k$ and observation equation $h_k$ respectively. This note focuses on modeling of the Markov transition noise $w_k$.

Several standard SSMs are widely used as benchmark simulation model for evaluation and comparison of recursive filters, typically such as the univariate nonstationary growth model given in [3, 4] and target tracking model given in [3, 5, 6]. For example, the state dynamic equations of the well-known univariate growth SSM and multi-dimensional target tracking SSM are given as follows, respectively

$$x_k = \frac{x_{k-1}}{2} + \frac{25x_{k-1}}{1+x_{k-1}^2} + 8\cos(1.2(k-1)) + w_k \tag{2}$$

$$x_k = \begin{bmatrix} 1 & t & 0 & 0 \\ 0 & 1 & 0 & 0 \\ 0 & 0 & 1 & t \\ 0 & 0 & 0 & 1 \end{bmatrix} x_{k-1} + \begin{bmatrix} t^2/2 & 0 \\ t & 0 \\ 0 & t^2/2 \\ 0 & t \end{bmatrix} \begin{bmatrix} w_{1,k} \\ w_{2,k} \end{bmatrix} \tag{3}$$

where for the target tracking SSM $x_k=[x_{1,k}, x_{2,k}, x_{3,k}, x_{4,k}]^T$, $[x_{1,k}, x_{3,k}]^T$ is the position while $[x_{2,k}, x_{4,k}]^T$ is the velocity at time $k$ and the sampling time $t=1$.

There are models reported with dependent noise processes [5] or unknown parameters [6] and here I concern more generally about the dependence of transition noise $w_k$ on the duration of simulation iterations. Straight to the point, the dynamic characteristic of $w_k$ may be undermined in the process of discretization, since the simulation runs indeed according to recursive iterations $k$ whose duration is always treated as 'one unit' regardless of how long it really is in the real world. This generates a gap between the practice and the simulation.

For example, faster iterating frequency (shorter sampling period) generally accompanies with relatively smaller transition noise and lower iterating frequency (longer sampling period) corresponds to bigger interval noise in practice. Further, sample period often affects sensitivity to error and stability of a process. Therefore, the transition noise $w_k$ should match the sampling period of the filter on the specified model. That is to say, the state transition noise $w_k$ is in turn depending highly on the real-time performance of the filter on the simulation model. To compare filters with different processing speeds under the same model, it is necessary to use different transition noise with respect to each filter's processing speed separately. Otherwise, the simulation results can be far from the truth.

Unfortunately, this is widely overlooked in the community for evaluation/comparison of discrete filters such as Kalman filters, particle filters, etc. On the contrary, standard simulation models are used to test filters by using the same transition noise $w_k$ regardless of their real intervals, e.g. $w_k$ is always set to be a zero-mean Gaussian white noise with the same covariance 10 in the univariate non-stationary growth model (2), see e.g. [3, 4] and the process noise $\{w_{1,k}\}$, $\{w_{2,k}\}$ are mutually independent zero-mean Gaussian white noise with respective standard deviation $\sigma_{v1}=1$ and $\sigma_{v2}=0.1$, see [5, 6]. In them, the estimation accuracy is compared between filters with different speeds on the same simulation model with the same simulation noise. They are not a special case, but quit common in considerable publications (including the author's work [3]). These pure simulations may be beyond reproach. But if considered in practice, their simulation results say nothing since they only hold for systems in which the state transition noise is independent of the time and is constant.

This is already bad enough, but matters get worse. It may give the implication of using over-complicated strategies to improve the filtering accuracy which may work in the simulation that use the same parameters but not, at least not so good, in practice. To illustrate this, consider one case involved in particle filters. In case of fixed transition noise, the more particles to use, the better estimation accuracy to obtain. This, however, cannot be guaranteed at all in practice since further increasing of the number of particles on the basis of a large number already will not improve the estimation accuracy much, but will obviously increase the sampling interval and thereby the transition noise. This probably will reduce the estimation accuracy more than it can improve, since clearly it is not always the case that the more particles, the better accuracy. However, this fact conflicts with the simulations in which the transition noise is constant and independent of the iterating speed. In this fashion, the 'excellent' performance reported of various advanced filters under the same state transition noise in simulation is much suspicious.

*Conclusion*

This note points out a common unfair treatment of the transition noise in the discrete simulation modeling for recursive filters. The state transition noise that inherently depends on the sampling interval should be set to match the sampling period of the specified filter. Different filters should use different transition noise correspondingly for simulation. Otherwise, the simulation result can be far from the truth. This, in my view, deserves high attention from designers and users of all simulation models for discrete filtering. It seems hard to find a simple solution for this problem except the discrete simulation modeling makes some big change.

*Acknowledgment*

Tiancheng Li would like to thank Professor Yu-Chi (Larry) Ho with Harvard University, USA for his discussion (dozens of email feedback) and encourage on this note. The note has also received positive feedback from professor Djuric with New York Stony Brook University, USA and professor Gustafsson with Linköping University, Sweden.

# References

[1]  Crisan, D. & Rozovskii, B. (2011) The Oxford Handbook on Nonlinear Filtering. Oxford: Oxford University Press.

[2] James, M. R., Krishnamurthy, V. & Le Gland, F. (1996) Time Discretization of Continuous-Time Filters and Smoothers for HMM Parameter Estimation. *IEEE Transactions on Information Theory*, 42(2), 593-605.

[3] Li, T., Sattar, T. P., Sun, S. (2012), Deterministic resampling: unbiased sampling to avoid sample impoverishment in particle filters, *Signal Processing*, 92(7), pp. 1637-1645.

[4] Bar-shalom, Y. & Fortmann, T.E. (1988) Tracking and Data Association. San Diego, CA: Academic.

[5] Saha, S. & Gustafsson, F. (2012) Particle filtering with dependent noise processes. *IEEE Transactions on Signal Processing*, 60(9), 4497-4508.

[6] Punithakumar, K., Kirubarajan,T. & Sinha, A. (2008) Multiple-model probability hypothesis density filter for tracking maneuvering targets. *IEEE Transactions on Aerospace and Electronic Systems,* 44(1), 87-98.

[7] Schön, T. B., Wills, A. & Ninness, B. (2011) System identification of nonlinear state-space models. *Automatica*, 47(1), 39-49.